\documentclass{article}

\usepackage{arxiv}
\usepackage[utf8]{inputenc} 
\usepackage[T1]{fontenc}    
\usepackage{hyperref}       
\usepackage{url}            
\usepackage{booktabs}       
\usepackage{amsfonts}       
\usepackage{nicefrac}       
\usepackage{microtype}      
\usepackage{lipsum}
\usepackage{graphicx}
\usepackage{float}
\usepackage[inline]{enumitem}
\usepackage{graphbox}
\usepackage{tikz}
\usepackage{listings}
\usepackage{pythonhighlight}
\usepackage{tabularx}
\usepackage{grffile}
\usepackage{amsmath,mathpazo}
\usepackage{esvect}
\usepackage{algorithm}
\usepackage{adjustbox}
\usepackage{lscape}
\usepackage{wrapfig}
\usepackage{mathtools}
\usepackage{enumitem}
\usepackage{bm}
\usepackage{lscape}
\usepackage{subcaption} 
\usepackage[noend]{algpseudocode}
\makeatletter
\usepackage{geometry}
\usepackage{environ}
\usepackage{color,soul}
\usepackage{setspace}
\usepackage{titlesec}

\newcounter{myalgctr}

\newcommand*{\affaddr}[1]{#1} 

\newcommand*{\email}[1]{\texttt{#1}}

\numberwithin{myalgctr}{section}

\title{Length Scale Control in Topology Optimization using Fourier Enhanced Neural Networks}

\author{%
Aaditya Chandrasekhar,   Krishnan Suresh \\
\affaddr{Department of Mechanical Engineering, University of Wisconsin-Madison}\\
\email{\{achandrasek3, ksuresh\}@wisc.edu%
}}

\begin{document}
\maketitle
\begin{abstract}
Length scale control is  imposed in topology optimization (TO)  to make designs amenable to manufacturing and other functional requirements.  \textcolor {black} {Broadly, there are two types of length-scale control in TO: \emph {exact} and \emph {approximate}. While the former is desirable, its implementation can be difficult, and  is computationally expensive. Approximate length scale control is therefore preferred, and is often sufficient for early stages of design. }

In this paper we propose an \textcolor {black} {approximate} length scale control strategy for TO, by extending a recently proposed density-based TO formulation using neural networks (TOuNN). Specifically, we enhance TOuNN with a Fourier space projection, to  control the  minimum and/or maximum length scales. The proposed method does not involve additional constraints,  and the sensitivity computations are automated \textcolor{black}{by expressing the computations in an end-end differentiable fashion using the neural net's library.} The proposed method is illustrated through several numerical experiments \textcolor {black} {for single and multi-material designs.}
\end{abstract}

\section{Introduction}
\label{sec:introduction}

Perhaps the most influential legacy of late Prof. Herb Voelcker is his work on the representations of solids \cite{Requicha83}. In this paper, we investigate representing solids through a concatenation of sinusoidal functions and  neural network's activation functions, and exploiting this representation for length-scale control in topology optimization.

Topology optimization (TO) \cite{Bendsoe2003} is a design method which distributes material within a domain to minimize an objective function, subject to constraints.  Various constraints are often imposed to make the design amenable to manufacturing and other functional requirements. One such set of constraints is to limit the maximum and minimum length scale of structural members. Imposing  a maximum length $l_{max}$ will lead to  desriable beam-like structures, while imposing a minimum length  $l_{min}$ avoids extremely thin non-manufacturable members.

Several methods have been proposed to impose length scale control in TO (see Section \ref{sec:literatureReview}). In this paper we propose length scale control by extending a density-based TO formulation using neural networks (TOuNN)  proposed in \cite{ChandrasekharTOuNN2020}, \cite{ChandrasekharMMTOuNN2020}. Specifically, in Section \ref{sec:method}, we enhance TOuNN with a Fourier space projection, leading to an approximate control on both the minimum and maximum length scales. The proposed method is illustrated through numerous experiments in Section \ref{sec:results}. Open research challenges, opportunities and conclusions are summarized in Section \ref{sec:conclusion}.

\section{Literature Review}
\label{sec:literatureReview}

Popular TO methods today include density based methods  (\cite{Bendsoe2003, sigmund2001Code99}), level-set methods \cite{Sethian2000}, and topological sensitivity methods (\cite{Suresh199Code, Deng2015,suresh2013efficient}). For a comprehensive review, see \cite{Rozvany2009,deaton2014survey}. Among these, density based methods, with solid isotropic material with penalization (SIMP) as the material model, is perhaps the most popular, and it serves as a basis for this paper.

\subsection{Topology Optimization using Mesh-Based SIMP}
\label{sec:litRev_TOSIMP}
\begin{figure}[h!]
	\begin{center}
		\includegraphics[scale=0.15,trim={1200 3150 0 0},clip]{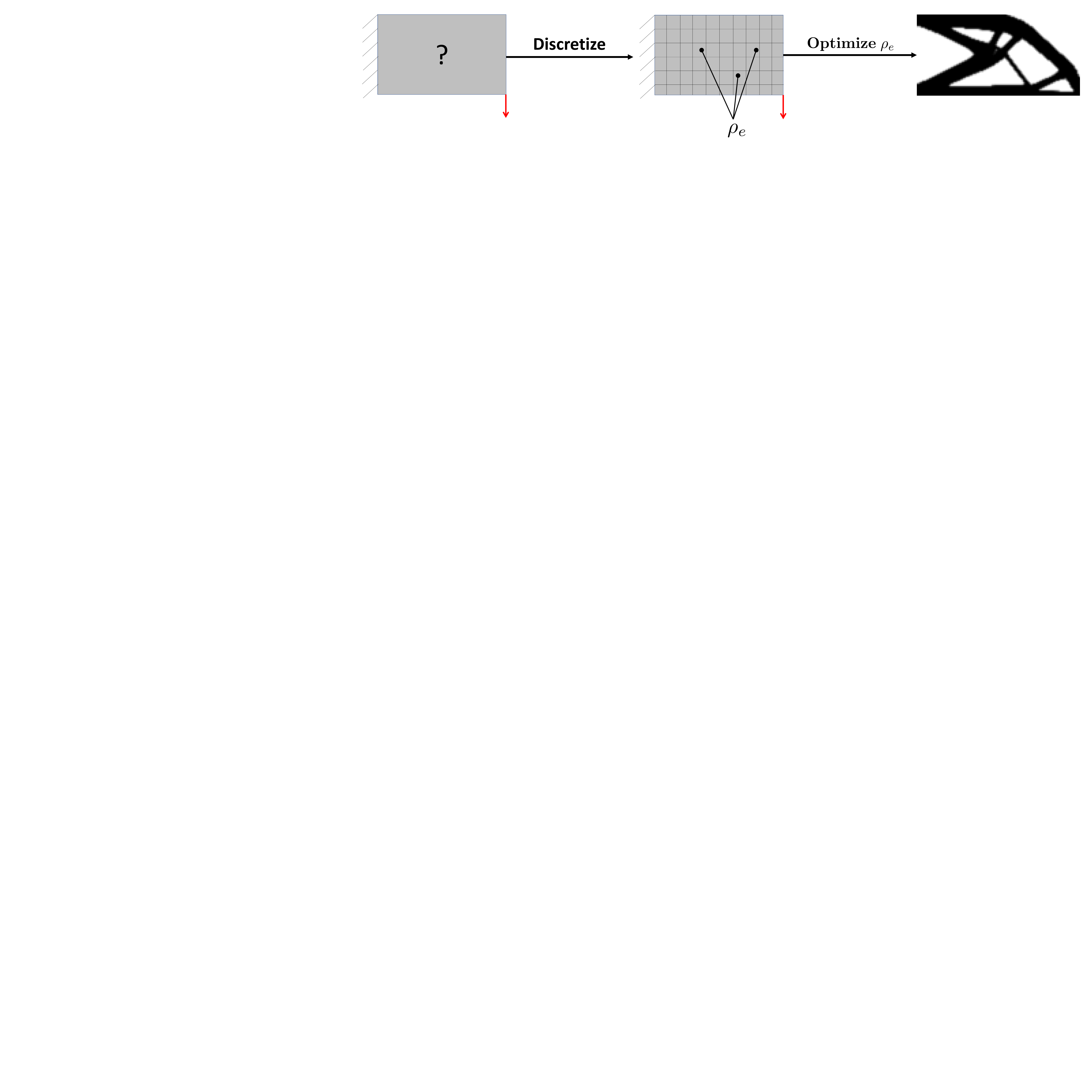}%
		\caption{Classic TO using mesh-based SIMP. }
		\label{fig:ClassicSIMPFlowchart}
	\end{center}
\end{figure}

 Consider the generic TO problem illustrated in Figure \ref{fig:ClassicSIMPFlowchart}. In density-methods, one defines a pseudo-density $\rho(\bm{x}) \in (0,1]$ over the domain, where  $\rho$ field is typically represented using an underlying mesh. Then, by relying on standard SIMP material penalization of the form $ E = E_0 \rho^p $, the field is optimized using optimality criteria \cite{Bendsoe2003} or MMA \cite{svanberg1987MMA}, resulting in the desired topology, as illustrated in Figure \ref{fig:ClassicSIMPFlowchart}.

Formally, the above TO problem can be posed as 
\textcolor{black}{
\begin{subequations}
	\label{eqnGenericMultiscaleOptimization}
	\begin{align}
	& \underset{\boldsymbol{\rho}}{\text{minimize}}
	& 		&\boldsymbol{u}^\mathsf{T}\boldsymbol{K}(\boldsymbol{\rho})\boldsymbol{u}\label{eqnObj_original}\\
	& \text{subject to}
	& & \boldsymbol{K}(\boldsymbol{\rho})\boldsymbol{u} = \boldsymbol{f}\label{eqn:GoverningEqn_original}\\
	& & & \sum_e \rho_e v_e \leq V^*\label{eqn:genericConstraint_original}
	\end{align}
\end{subequations}
}
where $\boldsymbol{u}$ is the displacement field, $\boldsymbol{K}$ is the finite element stiffness matrix, $\boldsymbol{f}$ is the applied force, $\rho_e$ is the density associated with element $e$, $v_e$ is the volume of the element, and $V^*$ is the prescribed volume.

\subsection{Length Scale Control in Topology Optimization}
\label{sec:litrev_featureSizeTO}

Length-scale control \textcolor{black}{, i.e., controlling feature thickness orthogonal to the local orientation,} is often imposed in TO to make the design amenable to manufacturing and other functional requirements; see \cite{Lazarov2016LengthScaleReview} for a review. Broadly, there are two types of length-scale control: \emph {exact} and \emph {approximate}.

\emph {Exact} length control implies that all features must lie precisely within the length scale specified. While this may be desirable during the final stages of design, it  is difficult to implement,  computationally demanding and often an overkill during early stages of design. The skeletal based minimum and maximum length scale control presented in \cite{Zhang2014SkeletonLengthScale} is an exact length-scale strategy; practical challenges in imposing exact length-scale control were reported in this work.  Further, exact length scale  can contradict imposed volume fraction constraint as explained later in Section \ref{sec:expts_minLengthScale}. Researchers have therefore turned towards approximate length scale control (see discussion below) since this is often sufficient for conceptual designs, AM infills,  micro-structural designs, etc. 

Several \emph {approximate} length scale controls methods have  been proposed. Filtering schemes to eliminate mesh dependency and indirectly control minimum length scale \cite{sigmund1997design}  is one the simplest approximate length-scale control method. In \cite{guest2004MinLengthScaleProjection}, a projection filter with tunable support was suggested to control the minimum length scale. However, these filters result in topologies with diffused boundaries. A scheme for imposing maximum length scale was developed in \cite{Guest2009MaxLengthScale}. In this approach, the radius of a circular test region determines the maximum allowable member size. However, the large number of nonlinear constraints  can pose challenges. Therefore, the constraints were aggregated using a p-norm form in \cite{Wu2017, Wu2017ShellInfill}, within the context of infill optimization.  The authors of \cite{Fernandez2020} recently introduced constraints  for minimum and maximum member size, minimum cavity size, and minimum separation distance.   Apart from requiring a large number of elements to obtain crisp boundaries, the computation and storage of these filters can also be expensive.   While most of the filters discussed above directly apply on the design variables, the readers are referred to \cite{sigmund2012sensitivityFiltering} for a discussion on  filtering applied to sensitivities, and to \cite{sigmund2009manufacturing} for imposing manufacturing constraints using filters in TO.

Imposing approximate length scale control in the frequency domain has been suggested in \cite{Lazarov2017FourierLengthScale} where a band pass filter was constructed to impose both maximum and minimum length scale controls. While this avoids  additional constraints,  the proposed method was restricted to rectangular domain, with non-rectangular domains requiring padding. Other techniques using morphology filters \cite{Sigmund2007MorphologyFilter},  geometric constraints \cite{Zhou2015MinLengthScaleGeomCons} robust length scale control \cite{Schevenels2011} have also been proposed.  Physics-based length-scale strategies include an implicit strain energy approach  \cite{zhu2012LevelSetLengthScale}, and a quadratic energy functional \cite{chen2008shape}.

The work described in this paper relies on a Fourier projection for length-scale control. \textcolor{black}{ The concept of using Fourier basis functions is not new. It was first explored in \cite {gomes2006application} within a level-set formulation. It was later extended to density-methods in \cite{Lazarov2017FourierLengthScale} via explicit  band-pass filters that control both the minimum and maximum length scales. However, the authors noted that \cite{Lazarov2017FourierLengthScale} \emph {" ... it is applicable to regular rectangular domains only. Irregular domains can be extended to rectangular ones by padding the design field with zeros, which can increase the computational cost associated with the filter."} This was then generalized to non-rectangular domains in \cite{white2018TOFourier}, where the authors noted that   \emph {"...[Fourier projection] decouples the geometry description,
and hence the decision variables, from the finite element
mesh."} }

\textcolor {black} {In this paper, we once again use Fourier projection for length-scale control. However, instead of directly using the Fourier coefficients as design variables, a neural network (NN) is used here to control these coefficients. A few advantages of this approach are: (1) the framework can be easily extended to, say, multiple-materials, without increasing the number of design variables, (2) one can extract high resolution topology, without increasing the computational cost, and (3) one can leverage the NN's built-in optimizer and back-propagation capabilities for automated sensitivity computations. } 

\subsection{ Topology Optimization using Neural Networks}
\label{sec:litRev_TOuNN}

\begin{figure}
	\begin{center}
		\includegraphics[scale=0.8,trim={50 535 20 20},clip]{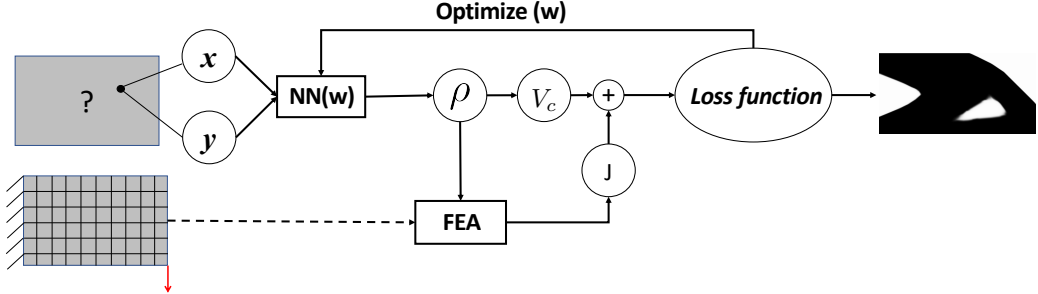}%
		\caption{TOuNN framework \cite{ChandrasekharTOuNN2020}. }
		\label{fig:TOuNNFlowchart}
	\end{center}
\end{figure}

The proposed method builds upon the TOuNN framework  \cite{ChandrasekharTOuNN2020} that we briefly review. Consider once again a generic TO problem in Fig \ref {fig:TOuNNFlowchart}. As in mesh based density methods, we define a pseudo-density field $\rho$  \textcolor{black} {defined at all points $(x,y)$ within the domain}. However, instead of representing $\rho$  using the mesh, it is represented here using a fully-connected neural network (NN). In other words, $\rho$ is spanned by NN's activation functions and controlled by the weights $\bm{w}$ within the NN. Then, by relying on SIMP material model, the density is optimized using the NN’s built-in optimizer  to respect the imposed volume constraint $V_c$ , while minimizing the compliance $J$, to result in the desired topology; see Figure \ref{fig:TOuNNFlowchart}. \textcolor {black} {Thus, as noted in \cite{white2018TOFourier}, the representation scheme for the density field is independent of the mesh. However, the representational parameters, i.e., the weights, indirectly depend on the mesh, through the discrete solution to the state equation, as described in the remainder of the paper.}

\begin{figure}
	\begin{center}
		\includegraphics[scale=0.10,trim={60 2450 0 0},clip]{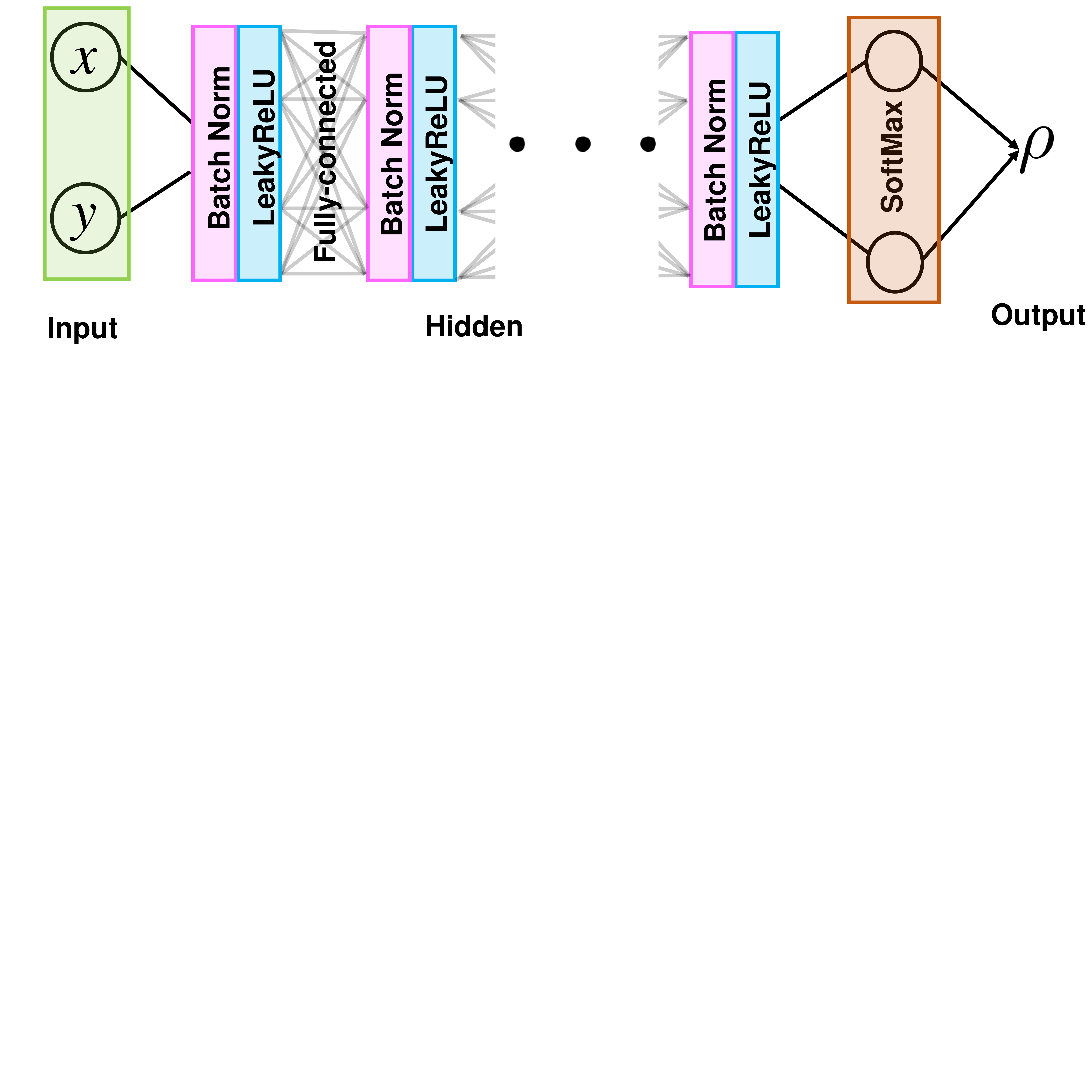}%
		\caption{The architecture of the neural net in the simple TOuNN framework \cite{ChandrasekharTOuNN2020}. }
		\label{fig:classicTOuNNNetwork}
	\end{center}
\end{figure}

As illustrated in Figure \ref{fig:classicTOuNNNetwork}, the NN typically consists of several hidden layers (depth); each layer may consist of several activation functions such as LeakyReLU \cite{Goodfellow2016_deepLearning}, \cite{lu2019dyingReLU} coupled with batch normalization \cite{Ioffe2015BatchNorm}. By varying the height and depth, one can increase the representational capacity of the NN. The final layer is a softMax function that scales the output to values between 0 and 1. 

Formally, in TOuNN, the topology optimization problem is posed as:
\textcolor{black}{
\begin{subequations}
	\label{eqnDensityNN}
	\begin{align}
	& \underset{\boldsymbol{w}}{\text{minimize}}
	& &\boldsymbol{u}^\mathsf{T}\boldsymbol{K}(\boldsymbol{w})\boldsymbol{u}\label{eqnObj}\\
	& \text{subject to}
	& & \boldsymbol{K}(\boldsymbol{w})\boldsymbol{u} = \boldsymbol{f}\label{eqn:GoverningEqn}\\
	& & & \sum_e \rho_e (\boldsymbol{w}) v_e \leq V^*\label{eqn:genericConstraint}
	\end{align}
\end{subequations}
}
where $\boldsymbol{w}$ are the weights associated with the NN. \textcolor{black}{Note that TOuNN  relies on SIMP penalization of the form $ E = E_0 \rho^p $ to drive the density field towards 0/1.}

The key characteristics of TOuNN are \cite{ChandrasekharTOuNN2020}: 
\begin{enumerate}
\item  The design variables are the NN weights $\boldsymbol{w}$.
\item The field $\rho(\bm{x})$ is infinitely differentiable everywhere in the domain.
\item The sensitivities are computed analytically using NN’s back-propagation.
\item  Optimization is carried out using NN's built-in optimizer.
\end{enumerate}

\section{Proposed Method}
\label{sec:method}

\begin{figure}
	\begin{center}
		\includegraphics[scale=0.85,trim={100 490 10 20},clip]{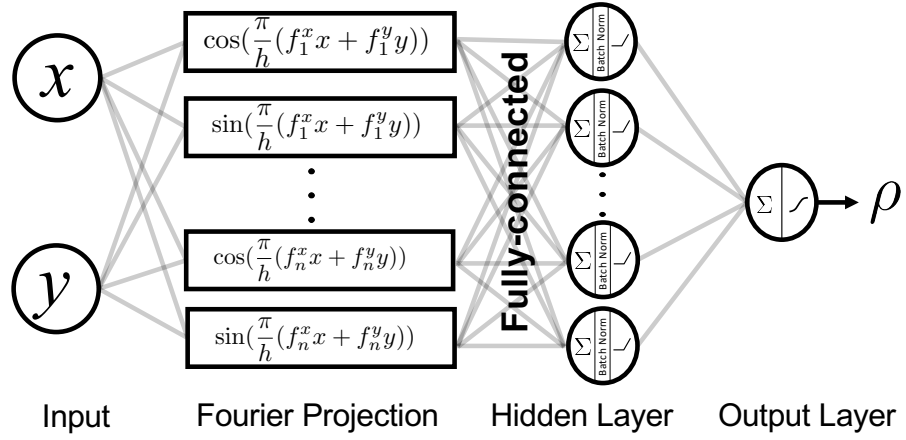}%
		\caption{The architecture of the proposed neural net for 2D problems. }
		\label{fig:fourierTOuNNNetwork}
	\end{center}
\end{figure}

In theory, one can capture $\textit{any}$ density field in TOuNN by simply increasing the number of weights and tuning them appropriately. However, for reasons discussed in \cite{Rahaman2019NNSpectralBias}, standard NNs fail to efficiently capture high frequencies. This has been  reported, for example, in capturing  volumetric density \cite{mildenhall2020nerf}, occupancy \cite{mescheder2019occupancy}, signed distances\cite{park2019deepsdf}. Analysis shows that the eigenvalue spectrum of these networks decay rapidly as a function of frequency \cite{tancik2020fourierNN}. To alleviate this problem, projecting the input spatial coordinates to a  Fourier space was proposed in \cite{tancik2020fourierNN}.  With this as motivation, the proposed architecture of the Fourier enhanced NN for TO is illustrated in Figure \ref{fig:fourierTOuNNNetwork}.

 The critical features of the proposed Fourier-TOuNN framework are:

\begin{enumerate}
\item \textbf{Input Layer}: As before, the input to the Fourier-NN is any point $\bm{x} = (x,y) \in \mathbf{R}^2$ in the domain.
\item \textbf{Fourier Projection Layer}: This layer is the focus of the current work. The $2$ dimensional Euclidean input  is transformed to a $2n_f$ dimensional Fourier space (see Section \ref {sec:FourierProjection}), where $n_f$ is the chosen number of frequencies. The range of frequencies is determined by the length scale controls:
\begin{equation}
f \in \big[ \frac{h}{l_{\text{max}} },  \frac{h}{l_{\text{min}}}\big]
\label{eq:fourierFrequencies}
\end{equation}

where $h$ is the mesh edge length. Further, in this paper, the frequencies are uniformly sampled within the specified interval.

\item \textbf{Hidden Layers}: The output from the projected Fourier space is piped into a single layer of  neurons (compared to multiple layers in Figure \ref{fig:classicTOuNNNetwork}). This hidden layer is important in that it introduces an essential non-linearity. As in TOuNN, each of the neurons is associated with an activation function such as leaky-rectified linear unit (LeakyReLU), i.e., $LR^{\epsilon} (x)=  max(0,x)  + \epsilon min(0,x)$.  Batch normalization is used for regularization. 

\item \textbf{Output Layer}: The final layer consists of one neuron with a Sigmoid activation function, ensuring that the output density lies between 0 and 1.

\end{enumerate}

In summary the output density (ignoring batch normalization) can be expressed as:
\begin{equation}
\rho = \frac{1}{1+exp\bigg(-\sum\limits_j w_j LR^{\epsilon}_j\big(\sum\limits_{i = 1}^{n_f} w_{ij} \cos(\frac{\pi}{h}(f_i^x x + f_i^y y)) + \sum\limits_{i = n_f}^{2n_f} w_{ij} \sin(\frac{\pi}{h}(f_i^x x + f_i^y y)) \;\big) \bigg) }
\label{eq:NNDensity}
\end{equation}
where $ LR^{\epsilon}_j$ represents the LeakyReLU operator.

\subsection{A Simplified Scenario}
\label{sec:FourierProjection}
\begin{figure}
	\begin{center}
		\includegraphics[scale=1.0,trim={320 620 0 80},clip]{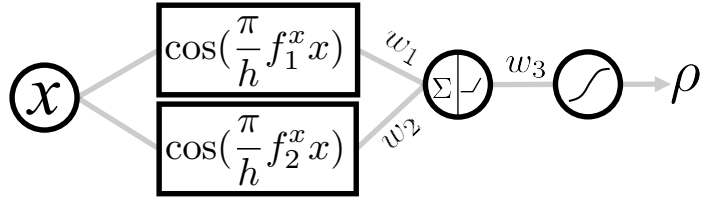}%
		\caption{Simplified representation of Neural Network with Fourier Projection. }
		\label{fig:simpleFourierNetwork}
	\end{center}
\end{figure}

To comprehend the transformations within the Fourier-TOuNN netowrk, consider a simplified version  in Figure \ref{fig:simpleFourierNetwork}, where we limit the input dimension to 1, the frequency terms are limited to 2, and only one LeakyReLU neuron is used. The output is a single Sigmoid function as before.  Then, the closed form expression for density simplifies  to:
\begin{equation}
\rho(x) = \frac{1}{ 1+ exp(-w_3 LR^{\epsilon}(w_1 \cos(\frac{\pi}{h} f_1^x x) + w_2 \cos( \frac{\pi}{h} f_2^x x)  ))}
\label{eq:simplifiedNNDensityExpression}
\end{equation}

\begin{figure}
	\begin{center}
		\includegraphics[scale=0.45,trim={75 0 0 0},clip]{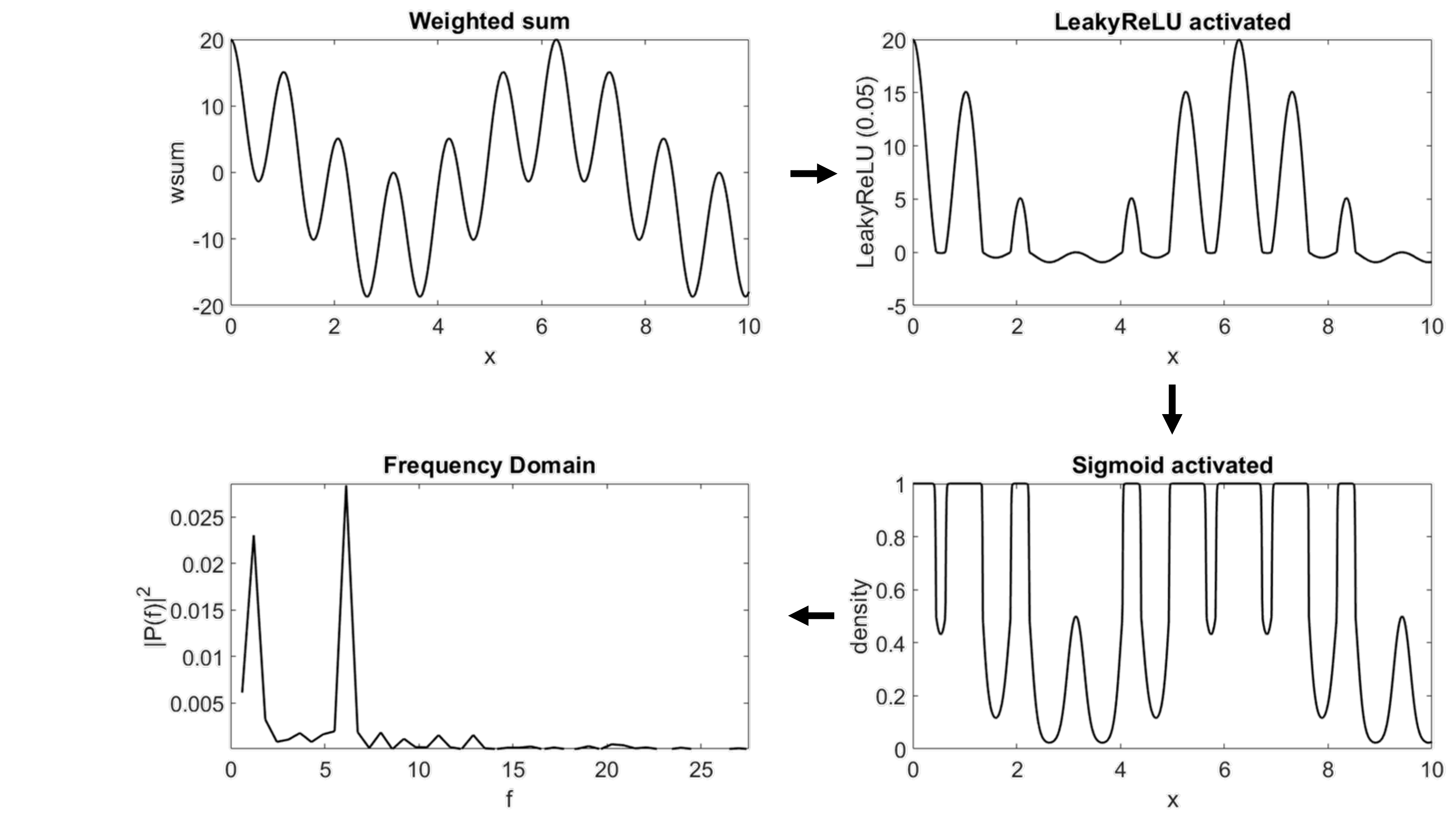}%
		\caption{Different stages of transformation in a Fourier-TOuNN. }
		\label{fig:theoreticalFunction}
	\end{center}
\end{figure}

Consider an instance where $w_1 = 8$, $w_2 = 10$, $w_3 = 4$, $f_1^x = 1$ and $f_2^x = 6$. Figure \ref{fig:theoreticalFunction} illustrates the different stages of the transformation. The plot on the top left is a simple linear combination of the two frequencies, while the plot on the top right is the output from the LeakyReLU function, which is essentially a linear transformation in the positive region, but that thresholds the negative valyes to $-\epsilon$. The Sigmoid (bottom right) further thresholds the output to $(0,1]$. The plot on the bottom left is the Fourier decomposition of the density field. As expected, the peaks are at the specified  frequencies of $1$ and $6$, but spread out. This spreading of the frequencies is discussed, for example, in \cite{tancik2020fourierNN}.  One can expect similar behavior in higher dimensions, and for a larger set of frequencies. \textcolor{black}{Note that the Sigmoid output can be anywhere in the range $[0,1]$. However, since we use SIMP as a material model within the proposed optimization method, intermediate densities are penalized, driving the output towards 0/1 \cite{du2015modified}. }

\subsection{Finite Element Analysis}
\label{sec:method_FEA}

Although the density field is represented using the NN, a finite element mesh is needed to capture the underlying physics. Here, we use regular node quad elements, and fast Cholesky factorization  based on the CVXOPT library \cite{andersen2013cvxopt}. During each iteration, the density at the center of each element is computed by the NN, and is provided to the FE solver. The FE solver computes the stiffness matrix for each element, and the assembled global stiffness matrix is then used to determine the  field $\bm{u}$ and the unscaled element compliances:
\begin{equation}
J_e = \{u_e\}^T[K]_0\{u_e\}
\label{eq:elemCompliance}
\end{equation}
The total compliance is given by:
\begin{equation}
\label{eq:totalCompliance}
J = \sum_e \rho_e^p J_e
\end{equation}
where $ p$ is the usual SIMP penalty parameter. 

\subsection{Loss Function}
\label{sec:method_lossFunc}
 While many optimization methods have been proposed for use within NN (see \cite{Schmidt2020}),  we use the Adam optimizer  \cite{Kingma2015ADAM}, a  gradient based optimization algorithm  implemented within PyTorch. Adam optimization is designed to minimize an unconstrained loss function. \textcolor{black}{In a volume constrained compliance minimization problem, it is reasonable to assume that the volume constraint in Eq. \ref{eqn:genericConstraint} is  active \cite{stolpe2010some} , i.e., it can  be treated as an equality constraint. This would allow the use of a simple penalty method of optimization as discussed below. Towards this end, } we convert the constrained minimization problem in Equation \ref{eqnDensityNN} into an unconstrained loss ${L}$ using a penalty formulation \cite{nocedal2006numericalOptimization}:

\begin{equation}
L(\bm{w}) = \frac{\sum\limits_e \rho_e^p J_e}{J_0} + \alpha \bigg(\frac{\sum\limits_e \rho_e v_e }{V^*} - 1 \bigg)^2
\label{eq:lossFunction}
\end{equation}
where $J_0$ is the initial compliance (for scaling), $p$ is the SIMP penalization parameter and $\alpha$ is the constraint penalty. As described in \cite{nocedal2006numericalOptimization}, starting from a
small positive value for the penalty parameter $\alpha$, a gradient driven step is taken to minimize the loss function. Then, the penalty parameter is increased and the process is repeated. Observe that, in the limit $\alpha \rightarrow \infty$, when the loss function is minimized, the equality constraint is satisfied and the objective is thereby minimized.

\subsection{Sensitivity Analysis}
\label{sec:sensAnalysis}

The Adam optimizer, being a first-order method, requires gradient information.  In order to compute the sensitivity of the loss function (that is being minimized) with respect to the design variables, note that: (1) the design variables are the weights of the NN  (instead of the element densities as in standard element-based TO), (2) the loss function comprises of the compliance and volume constraint through a penalty formulation.  

With these observations, the sensitivity of the loss function with respect to any design variable $w_i$ is given by:
\begin{equation}
\frac{\partial L}{\partial w_i} = \sum\limits_e \frac{\partial L}{\partial \rho_e} \frac{\partial \rho_e}{\partial w_i}
\end{equation}
The  term $\frac{\partial \rho_e}{\partial w_i}$ is the sensitivity of element density with respect to the weights.  Since the NN is nothing but a computational graph consisting of weighted sums and nonlinear activations,  automatic differentiation  \cite{baydin2017automatic} can be used to efficiently and accurately compute these terms.  In particular, we utilize the automatic differentiation  within PyTorch \cite{NEURIPS2019_9015_pyTorch} in our implementation.\\

\textcolor{black} {However, since the finite element solver is outside of the neural network, the sensitivity of compliance with respect to the density variables must be computed analytically.} Thus,  the sensitivity of the loss function with respect to element density, i.e., $\frac{\partial L}{\partial \rho_e}$, is obtained by differentiating Equation (\ref{eq:lossFunction}) and incorporating the self-adjoint term \cite{Bendsoe2003}, resulting in:

\begin{equation}
\label{eq:Sensitivity}
\frac{\partial L}{\partial w_i}  = \frac{-p}{J_0}\bigg(  \sum_e \rho_e^{p-1} J_e  \bigg)+ \frac{2 \alpha }{V^*} \bigg(\frac{\sum\limits_k \rho_k v_k}{V^*} - 1 \bigg)
\end{equation}
The two terms are combined within the framework to compute the desired  sensitivities.

\textcolor{black} {
\subsection{Extension to Multi-Material}}
\label{sec:method_FourierMMTO}
\textcolor{black} 
{ A key feature of the proposed framework is its versatility. Here, we show how the framework can be easily extended to multiple materials. In particular, we extend the multi-material framework proposed in \cite{ChandrasekharMMTOuNN2020} by adding length-scale control.}

\begin{figure}
	\begin{center}
		\includegraphics[scale=0.85,trim={0 0 0 0},clip]{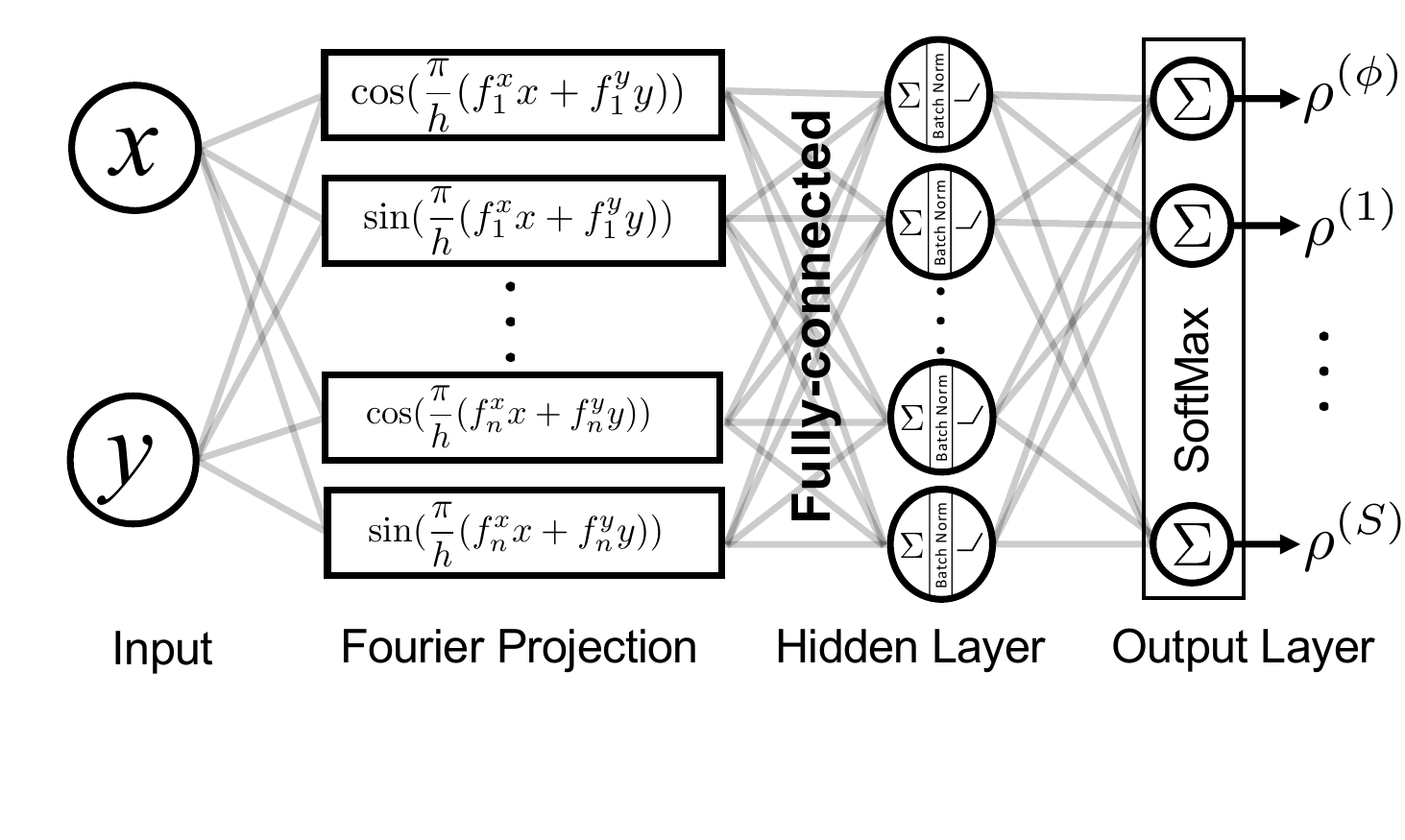}%
		\caption{\textcolor{black}{Length scale-control for multi-material  TO. }}
		\label{fig:MM_FourierTOuNN}
	\end{center}
\end{figure}

\textcolor{black}
{Recall from  Fig. \ref{fig:fourierTOuNNNetwork} that the NN for a single material had only one output neuron. To extend this to multiple materials, we simply  insert additional outputs as illustrated in Fig. \ref{fig:MM_FourierTOuNN}, corresponding to void $\rho^{(\phi)}$, and any of $S$ candidate materials, with corresponding mass densities $\rho^1, \ldots \rho^S$. The Softmax activation of the output layer ensures that $0 < \rho^{(i)} < 1$. Further, the Softmax construction also ensures that the partition of unity is satisfied, i.e., $\rho^{(\phi)} + \rho^{1} + \ldots \rho^{S} = 1$; please see \cite{ChandrasekharMMTOuNN2020} for a proof. The critical point is that one can use the same NN as was used for single material, i.e., the number of design variables remains almost unchanged, except for a few additional variables in the last layer. This is later substantiated  through numerical experiments. Finally, since the computation is end-to-end differentiable, one can avoid manual sensitivity derivation. Note that, for multi-material design, a net mass constraint must be imposed instead of a volume constraint; see \cite{ChandrasekharMMTOuNN2020}.
}

\section{Algorithm}
\label{sec:algorithm}
The optimization procedure is summarized in Algorithm \ref {alg:Fourier-TOuNN}. We will assume that the NN has been constructed with a desired number of layers, nodes per layer and activation functions. The NN is initialized using a Glorot initializer \cite{glorot2010understanding}. We discretize the domain into a mesh ($\Omega^h$) with desired number of elements.  Observe that mesh discretization is essential to capture the physics, but is not directly used to representat  the density field. The coordinates of the element centers are used as inputs to the NN during optimization. The  density at the center of element $e$ is denoted by $\rho_e$. The loss function is  calculated from these discrete values. Here, we use the continuation scheme where the parameter $p$ is incremented to avoid local minima \cite{Rojas2015Continuation}, \cite{Sigmund1998NumericalInstabTO}. The  process is then repeated until termination. The algorithm is set to terminate if the percentage of gray elements (elements with densities between 0.05 and .95) $\epsilon_g = N_{grey}/N_{total}$  is less than a prescribed value. \textcolor{black}{Here, a termination criteria of 0.0025 is used for $\epsilon_g$. Note that the method may not converge if the termination criteria is too small, say, less than 0.001. This is analogous to the practical termination criteria of 0.01  used in classic SIMP \cite{Andreassen2011_88LineCode} for the density change. }

\begin{figure}
	\centering
		\includegraphics[scale = 0.5,trim={0 0 0 0},clip]{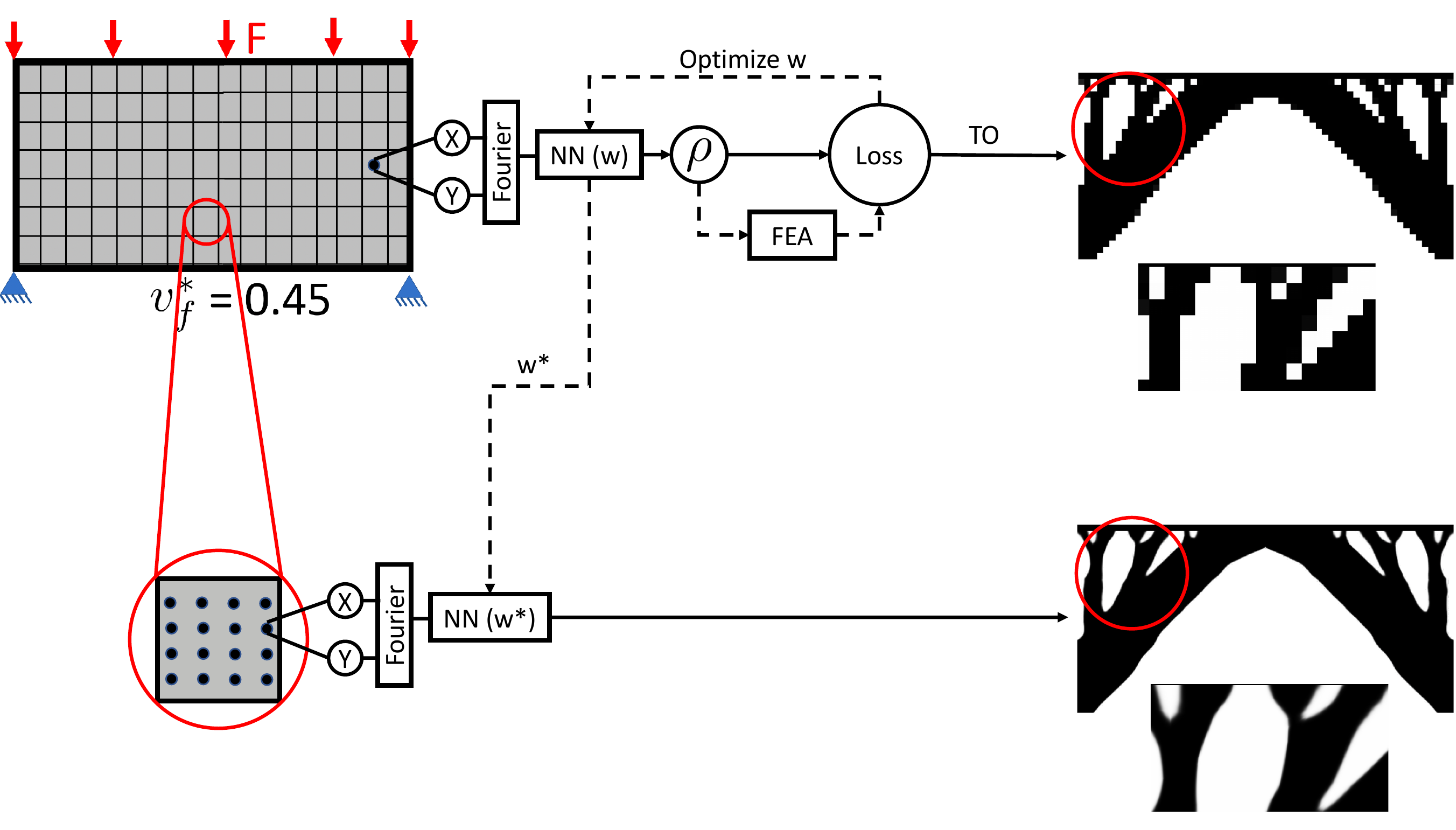}
	\caption{\textcolor{black}{ Extraction of high resolution topology from the NN.}}
	\label{fig:highResTopExtraction_distBeam}
\end{figure}

\textcolor{black}{A key feature of the proposed framework is that one can extract a high resolution topology at no optimization cost once the process is complete. Specifically, the optimized weights, i.e., $w^*$ , are used to sample the density field on a finer grid as illustrated in Fig. \ref{fig:highResTopExtraction_distBeam}.}

\begin{algorithm}
	\caption{Fourier-TOuNN}
	\label{alg:Fourier-TOuNN}
	\begin{algorithmic}[1]
		\Procedure{TopOpt}{NN, $\Omega^h, V^*, l_{min}, l_{max}$}
		\Comment{NN, discretized domain, desired vol, length scales}
		
		\State $\boldsymbol{x} = \{x_e,y_e\}_{e \in \Omega^h}$ or  $ \{x_e,y_e,z_e\}_{e \in \Omega^h}$ \; ; $\bm{x} \in \mathbf{R}^{n_e \times d}$\Comment{ center of elements in FE mesh; optimization input} \label{alg:domSampl}
		
		\State epoch = 0; $\alpha = \alpha_0$; $p = p_0$ \Comment{Penalty factor initialization}
		
		\State $J_0 \leftarrow FEA(\bm{\rho} = v_f^*, \Omega^h)$ \Comment {FEA with uniform gray}
		
		\State $f \sim \mathcal{U}(\frac{h}{l_{max}}, \frac{h}{l_{min}})$ \; ; $f \in \mathbf{R}^{d \times n_f}$ \Comment{Uniformly sampled frequencies} 
		
		\Repeat \Comment{Optimization}
		\State $\hat{\bm{x}} = \mathcal{F}(\bm{x} ; f)$ \; ; $\hat{\bm{x}} \in \mathbf{R}^{n_e \times 2 n_f}$ \Comment{Project from Euclidean to Fourier} 
		\State $\boldsymbol{\rho} = NN(\boldsymbol{\hat{x}} )$ \Comment{ Call NN to compute $\boldsymbol{\rho} $ at $\boldsymbol{p}$; (Forward propagation)}  \label{alg:densFrmNN}
		
		\State $J_e \leftarrow FEA(\bm{\rho},\Omega^h)$ \Comment{Solve FEA} \label{alg:fea}
		
		\State L = $\frac{\sum\limits_e \rho_e^p J_e}{J_0} + \alpha \bigg(\frac{\sum \limits_e \rho_e  v_e }{V^*} - 1\bigg)^2$ \Comment{ Loss function} \label{alg:loss}
		
		\State Compute $ \nabla L $\Comment{Sensitivity(Backward Propagation)}
		
		\State $ \boldsymbol{w}  \leftarrow \boldsymbol{w}  +  \Delta \boldsymbol{w} (\nabla L )  $ \Comment{Adam optimizer step}
		
		\State $\alpha   \leftarrow \text{min} (\alpha _{max}, \alpha + \Delta \alpha)$ \Comment {Increment $\alpha$} \label{alg:penaltyUpdate}
		
		\State $ p \leftarrow \text{min} (p_{max}, p + \Delta p)$ \Comment {Continuation } 
		
		\State $\text{epoch} \leftarrow \text{epoch}+1$
		\Until{ $\epsilon_g < \epsilon_g^*$ } \Comment{Check for convergence}
		
		\EndProcedure
	\end{algorithmic}
\end{algorithm}

\section{Numerical Experiments}
\label{sec:results}
In this section, we conduct several numerical experiments to illustrate the proposed algorithm.
The implementation is in Python. The default parameters are listed in Table \ref{table:defaultParameters}. \textcolor{black}{In this work, we use a continuation scheme for the SIMP penalty parameter $p$ driving it from $p=1$ at the start of optimization upto $p=8$.  This ensures stability of convergence \cite{Rojas2015Continuation, white2018TOFourier} as well as driving the final topology to a desired 0/1 state. The minimum number of iterations is set to 150, and a maximum to 500. After optimization a grid of $15 \times 15$ per element is used to extract a high resolution topology.}

\textcolor{black}{ The total number of design variables in the proposed framework can be derived based on the following argument. Observe that  $n_f$ cosine and sine components  are passed into the projection layer. Each of the terms is fully connected to  the $n_h$ neurons in the hidden layer. Further, each neuron in the hidden layer has a bias, and all the neurons in the hidden layer are mapped to one output neuron $(n_o)$. Thus the total number of design variables is given by, $n_f*2*n_h + n_h + n_h*n_o$. In our instance,  this results in a total of $150*2*20 + 20 + 20*1 = 6040$ design variables. The number of design variables is independent of the mesh discretization.}

\begin{table}
	\begin{center}
		\begin{tabular}{  r | l   }
			Parameter & Description and default value \\ \hline
			$E$, $\nu$ & Isotropic material constants Young's Modulus (E = 1) and Poisson's ratio ($\nu = 0.3$) \\
			$n_f$ & Number of frequencies in projection space = 150 \\
					NN & Fully connected feed-forward network with one layer (ReLU) with 20 Neurons.  \\
					$|\bm{w}|$ &   The number of design variables (weight and bias) is 6040 by default  \\
			$\alpha$ & Constraint penalty updated every iteration ($\alpha_0 = \Delta \alpha = 0.2$) \\
			$p$ & SIMP penalty updated every iteration ($p_0 = 1 \; ; \; \Delta p = 0.02 \; ; \; p_{max} = 8$) \\
			lr & Adam Optimizer learning rate 0.01 \\
			nelx, nely & Number of mesh elements of (60, 30)  along $x$ and $y$ respectively \\
			$\epsilon_g^*$ & Convergence criteria requiring fraction of gray elements be less than $\epsilon_g^* = 0.0025$ \\

		\end{tabular}
	\end{center}
	\caption{Default simulation parameters.}
	\label{table:defaultParameters}
\end{table}

All experiments were conducted on an Intel i7 - 6700 CPU @ 2.6 Ghz, equipped with 16 GB of RAM. Through the experiments, we investigate the following:
\begin{enumerate}
	\item $\textit{Benchmark Studies}$: First we consider several TO benchmark problems, and compare the topologies obtained through mesh-based SIMP, TOuNN, and proposed Fourier-TOuNN.

	\item $\textit{Comparison Studies}$: Next, we compare the computed topologies using Fourier-TOuNN against those obtained using other length scale strategies.
	
	\item $\textit{Convergence Study}$: The impact of the Fourier projection on the convergence of a distributed load problem is investigated.

	\item $\textit{Minimum Length-Scale Control}$: Here, we vary the minimum length scale ($l_{min}$) and study its effect on the topology, while keeping the maximum length scale ($l_{max}$) a constant.

	\item $\textit{Maximum Length-Scale Control}$: Similarly, we will vary the maximum length scale ($l_{max}$) and study its effect on the topology, while keeping the minimum length scale ($l_{min}$)  a constant.

	\item $\textit{Single Length-Scale Control}$: Next, we vary both $l_{min}$ and $l_{max}$, but keeping them equal, and study the impact on the topology.
	
	\item $\textit{Number of Frequency Samples}$: Next, we vary the number of frequency terms $n_f$ in the projected frequency space to understand its effects.
	
	\item$\textit{Depth of Neural Net}$: We vary the depth of NN to understand the nature of frequency spreading.
    
    \item $\textit{Multi-material design}$: \textcolor{black}{Finally, we present an extension of the proposed work to multi-material designs.}

\end{enumerate}
\subsection{Benchmark Studies}
\label{sec:expts_benchmark}

\begin{figure}
	\centering
	\includegraphics[scale = 0.55,trim={0 0 0 0},clip]{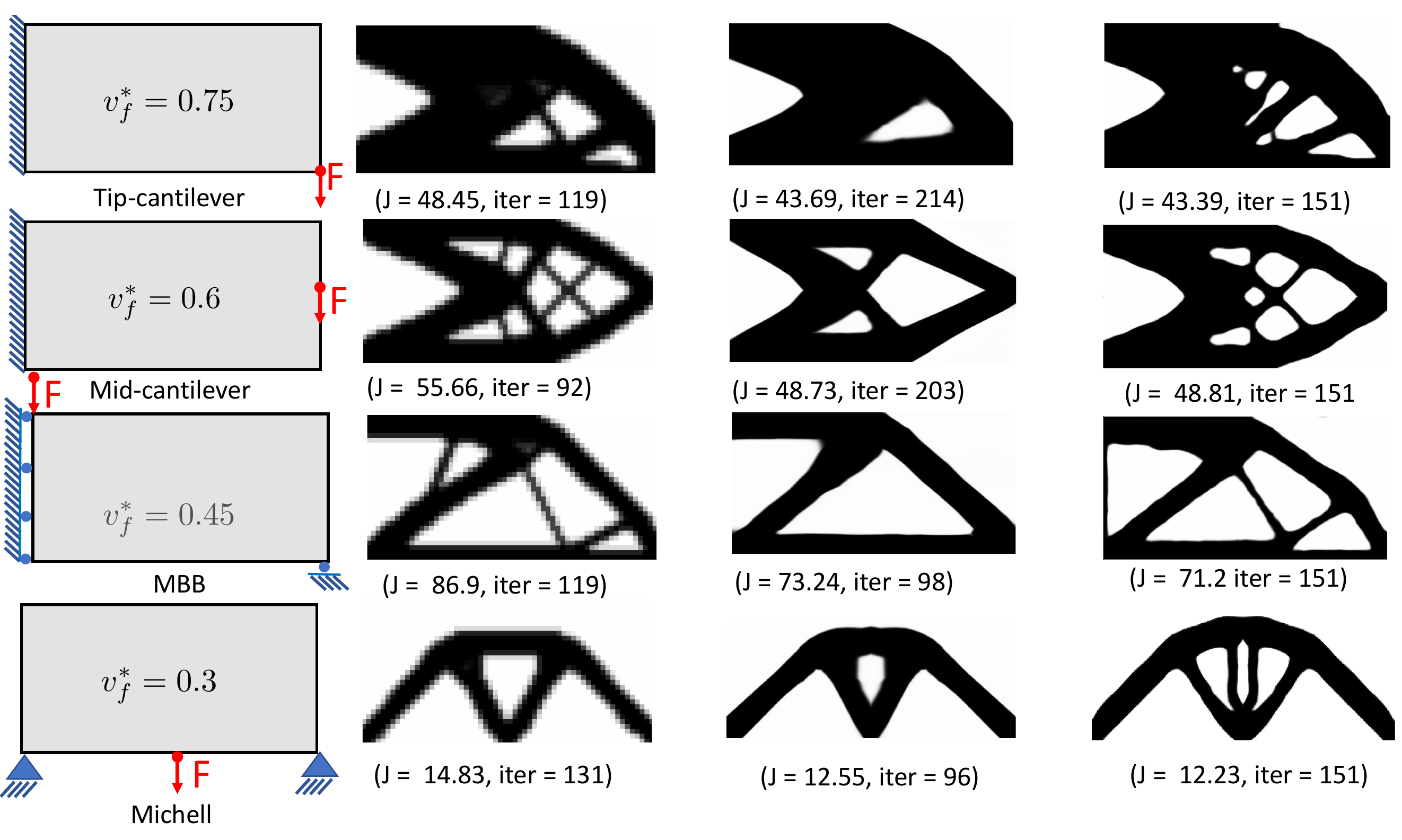}
	\caption{Comparison of topologies for various benchmark problems using SIMP \cite{Andreassen2011_88LineCode}, TOuNN \cite{ChandrasekharTOuNN2020} and proposed Fourier-TOuNN.}
	\label{fig:benchmark}
\end{figure}

First we consider  four benchmark topology optimization problems illustrated in the first column of Figure \ref{fig:benchmark}, together with the desired volume fraction. The default mesh size is $60 \times 30$. The second column illustrates the topologies obtained  using the popular 88-line implementation of SIMP \cite{Andreassen2011_88LineCode}, \textcolor{black}{with a filter radius of 1.4, that directly controls the minimum length scale, avoiding checker-board patterns \cite{Sigmund1998NumericalInstabTO}; no additional length-scale control  was imposed}.  The final compliance values and the number of finite element iterations are also listed. The third column illustrates the topologies obtained using TOuNN \cite{ChandrasekharTOuNN2020}, using a neural net size  of $4 \times 20$ on the same mesh, without length scale control. The fourth column are the topologies obtained through the proposed Fourier-TOuNN algorithm, with  $l_{min} = 6$ and $ l_{max} = 30$.  As one can  observe, the topologies computed through Fourier-TOuNN exhibit thin features, and the compliance values are lower (better) than both SIMP and TOuNN.  \textcolor{black}{We note that the plots in the third and fourth column are at a larger resolution. This stems from the fact that, after optimization, the density values are sampled on a much finer grid (than that used for FE) to extract fine geometry as explained in Fig. \ref{fig:highResTopExtraction_distBeam}; this comes with no additional computational cost. }The total computational time for the   examples in Figure \ref{fig:benchmark} varied between 4-6 seconds proportional to the number of FE iterations; this is consistent with other mesh-based implementations such as \cite{Andreassen2011_88LineCode}. Further, averaging across the examples,  we observed that, in Fourier-TOuNN, FE accounts for $41\%$, forward propagation accounts for $20\%$, backward propagation accounts for $27\%$, and the remaining $13\%$ is for miscellaneous operations. \textcolor{black}{This is in contrast to SIMP implementations where  FE accounts for a significant  portion of the computation, for instance about $96\%$ as reported in \cite{ferrari2020new}. Reducing the non-FE overhead in Fourier-TOuNN  via code refactoring and just-in-time compilations \cite{lam2015numba} remains to be explored. } 

\begin{figure}
	\centering
	\includegraphics[scale = 0.55,trim={0 0 0 0},clip]{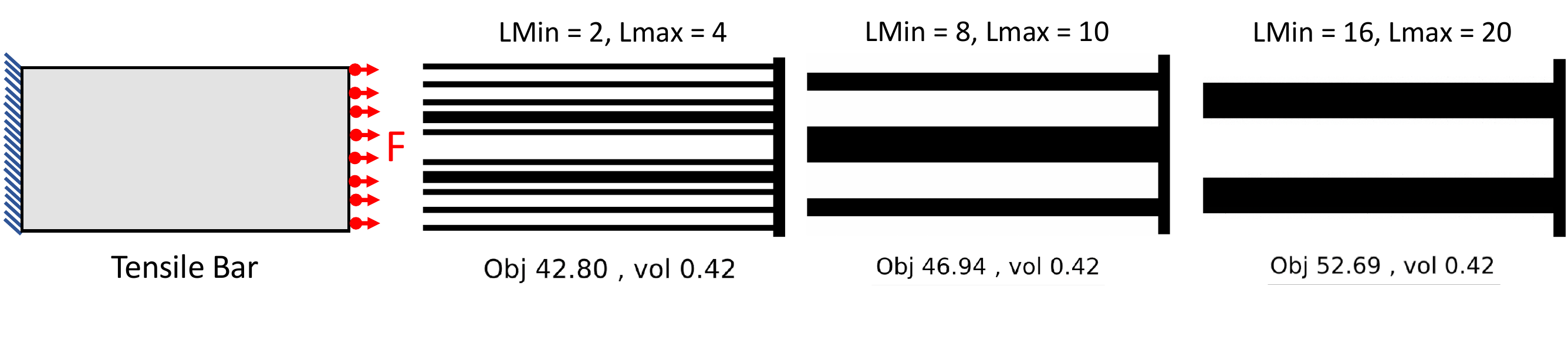}
	\caption{Tensile bar for varying length scale.}
	\label{fig:tensileBar}
\end{figure}

Next,  we consider a simple tensile bar problem, (over a mesh of size $60 \times 30$) as illustrated in Figure \ref{fig:tensileBar}. \textcolor {black} {The elements on the right edge are explicitly retained, while an extrude constraint along $x$ is imposed as discussed in \cite{ChandrasekharTOuNN2020}. Various length scale intervals are imposed  as illustrated; the topologies and compliances obtained are consistent with expectations. }

\subsection{Comparison Studies}
\label{sec:expts_comparision}

\begin{figure}
	\centering
	\includegraphics[scale = 0.55,trim={50 275 0 60},clip]{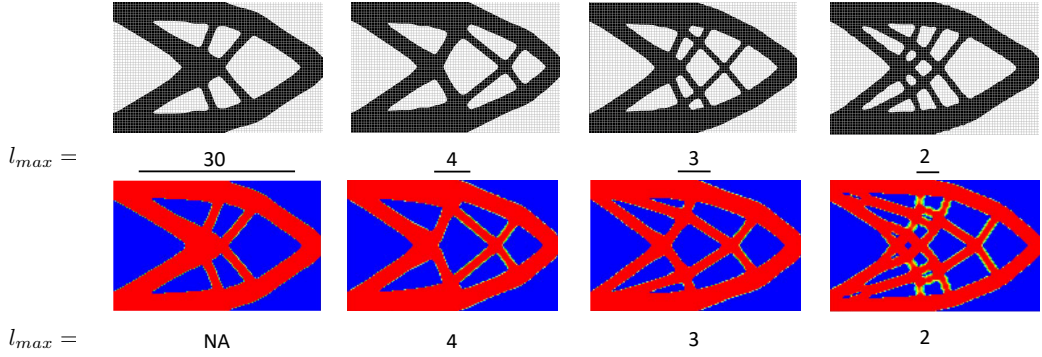}
	\caption{Benchmark comparison of designs from \cite{Guest2009MaxLengthScale} and proposed method. The maximum length scale is decreased (left to right) with  $l_{min} = 2$ (NA = Not Applicable). \textcolor{black}{The background grid on the top row reflects the FE mesh used.}}
	\label{fig:comparison_MidCantilever}
\end{figure}

We consider the length scale results presented in \cite{Guest2009MaxLengthScale} for a mid-cantilever beam with a mesh of $160 \times 100$. With a target volume fraction of $0.5$, the minimum length scale was kept constant at $l_{min} = 1$ while  $l_{max}$ was varied.   Figure \ref {fig:comparison_MidCantilever} illustrates the computed topologies using the proposed method (top row) and those reported in \cite{Guest2009MaxLengthScale} (bottom row). In both methods, as $l_{max}$ is decreased, one can observe more structural members beginning to appear. The compliance values are not compared here since it was not reported in \cite{Guest2009MaxLengthScale}.

\begin{figure}
	\centering
	\includegraphics[scale = 0.55,trim={0 390 0 80},clip]{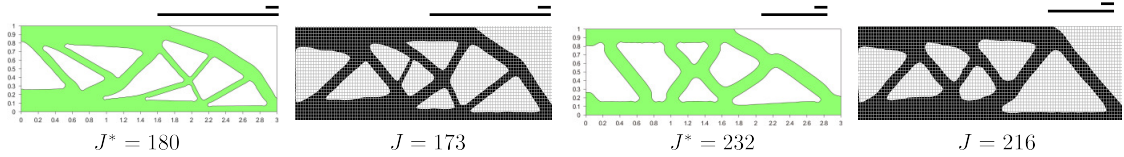}
	
	\caption{ (a) No length scale control \cite{Xia2015LengthScaleLevelSet}. (b) Relaxed length scale control in Fourier-TOuNN with $l_{min} = 3$  and $l_{max} = 100 $. (c) Tight length scale control \cite{Xia2015LengthScaleLevelSet} with $l_{min} = 5$  and $l_{max} = 8 $. (d) Tight length scale control in Fourier-TOuNN with $l_{min} = 5$  and $l_{max} = 8 $. \textcolor{black}{The background grid on the top row reflects the FE mesh used.}}
	\label{fig:comparison_MBB}
\end{figure}

Next we consider the results presented \cite{Xia2015LengthScaleLevelSet} for an MBB beam  with a mesh of $300 \times 100 $ elements, and a desired volume fraction of $V^* = 0.5$. The topology obtained without length scale control from \cite{Xia2015LengthScaleLevelSet} is illustrated in  Figure \ref{fig:comparison_MBB}a. Figure \ref{fig:comparison_MBB}b illustrates the topology obtained through Fourier-TOuNN with a relaxed length scale control of $l_{min} = 2$  and $l_{max} = 150 $. Figure \ref{fig:comparison_MBB} c is the reported topology \cite{Xia2015LengthScaleLevelSet}  with a tight and equivalent length control of  $l_{min} = 2$  and $l_{max} = 4$, while Figure \ref{fig:comparison_MBB}d illustrates the Fourier-TOuNN topology  with the same length control. We observe that length scale control increases the compliance as expected and more so, for exact length scale control.

\subsection{Convergence}
\label{sec:expts_convergence}
\begin{figure}
	\centering
		\includegraphics[scale = 0.5,trim={0 0 0 0},clip]{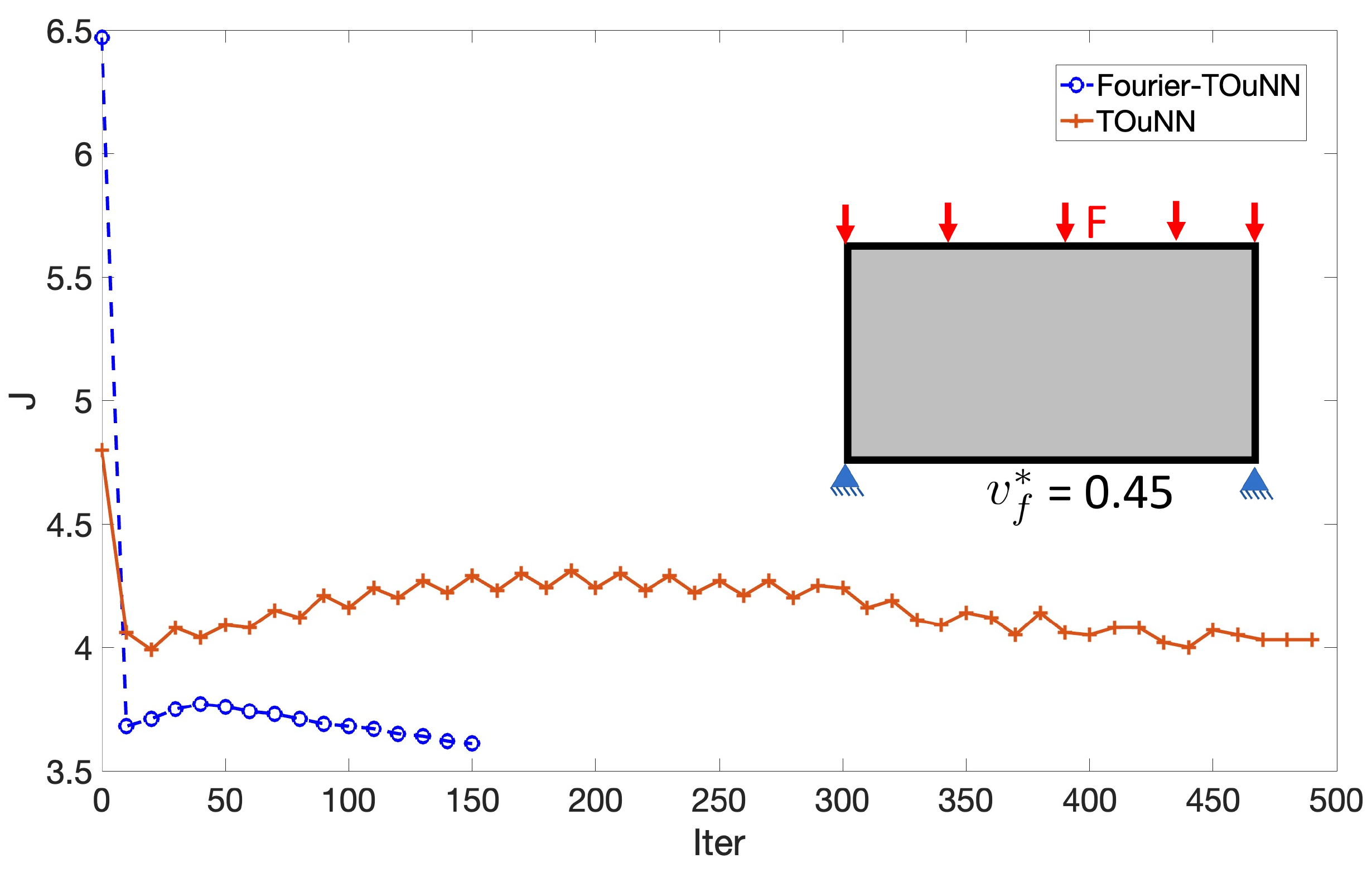}

	\caption{\textcolor{black}{ Convergence of compliance for distributed load at $V_f = 0.45$ using TOuNN \cite{ChandrasekharTOuNN2020} and  Fourier-TOuNN.}}
	\label{fig:convergence_distBeam_compliance}
\end{figure}

\begin{figure}
	\centering
		\includegraphics[scale = 0.5,trim={0 0 0 0},clip]{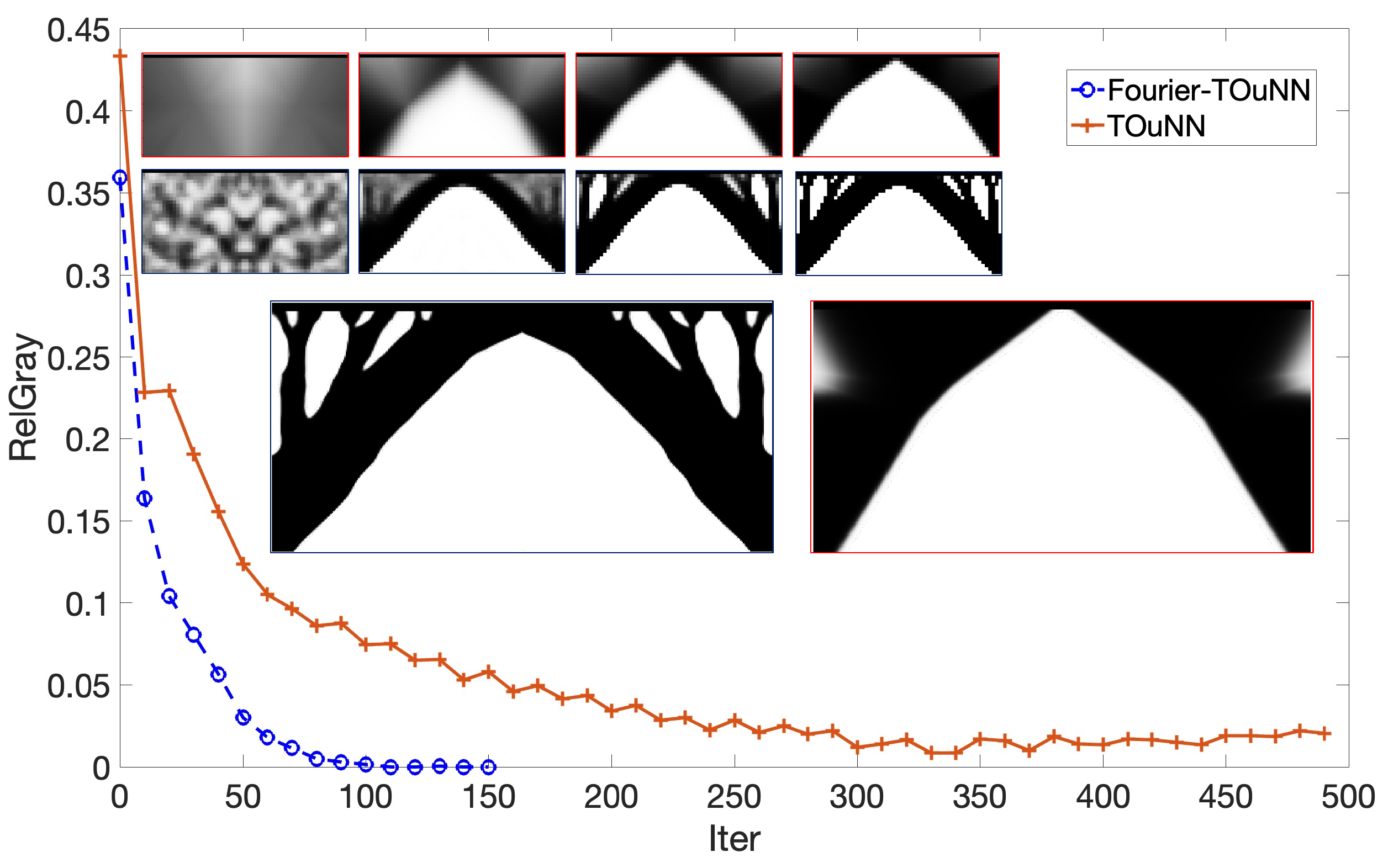}

	\caption{ \textcolor{black} {Progression} of relative amount of gray elements for distributed load at $V_f = 0.45$ using TOuNN \cite{ChandrasekharTOuNN2020} and  Fourier-TOuNN.}
	\label{fig:convergence_distBeam_gray}
\end{figure}

The distributed load problem  in Figure \ref{fig:convergence_distBeam_compliance} (inset) is considered next since it necessary entails thin features \cite{Mezzadri2018Support} near the loaded edge.  This problem is solved using two different methods: TOuNN \cite{ChandrasekharTOuNN2020} (with no length scale control), and with the proposed Fourier-TOuNN method with length scale control ($l_{max} = 30, l_{min} = 4$). \textcolor{black}{The elements at the top layer are retained explicitly.} \textcolor{black}{The difference in compliance convergence between TOuNN and Fourier-TOuNN is illustrated in  Figure \ref{fig:convergence_distBeam_compliance}, while Figure \ref{fig:convergence_distBeam_gray} illustrates the progression of the fraction of gray elements ($\epsilon_g = N_{gray}/N_{total}$)}.  We observe that TOuNN  is unable to resolve the topology due to the limitations of NN. However, the proposed Fourier-TOuNN  rapidly converges to a topology exhibiting thin features.

The poor convergence of TOuNN is consistent with the analysis in   \cite{tancik2020fourierNN} \textit{"for a conventional NN,  the eigenvalues of the Neural Tangent Kernel (NTK) decay rapidly. This results in extremely slow convergence to the high frequency components of the target function,  to the point where standard NNs are effectively unable to learn these components"}. On the other hand, the addition of the Fourier projection leads to rapid convergence.

\subsection{Minimum Length-Scale Control}
\label{sec:expts_minLengthScale}

\begin{figure}
	\begin{center}
		\includegraphics[scale=0.65,trim={70 120 0 0},clip]{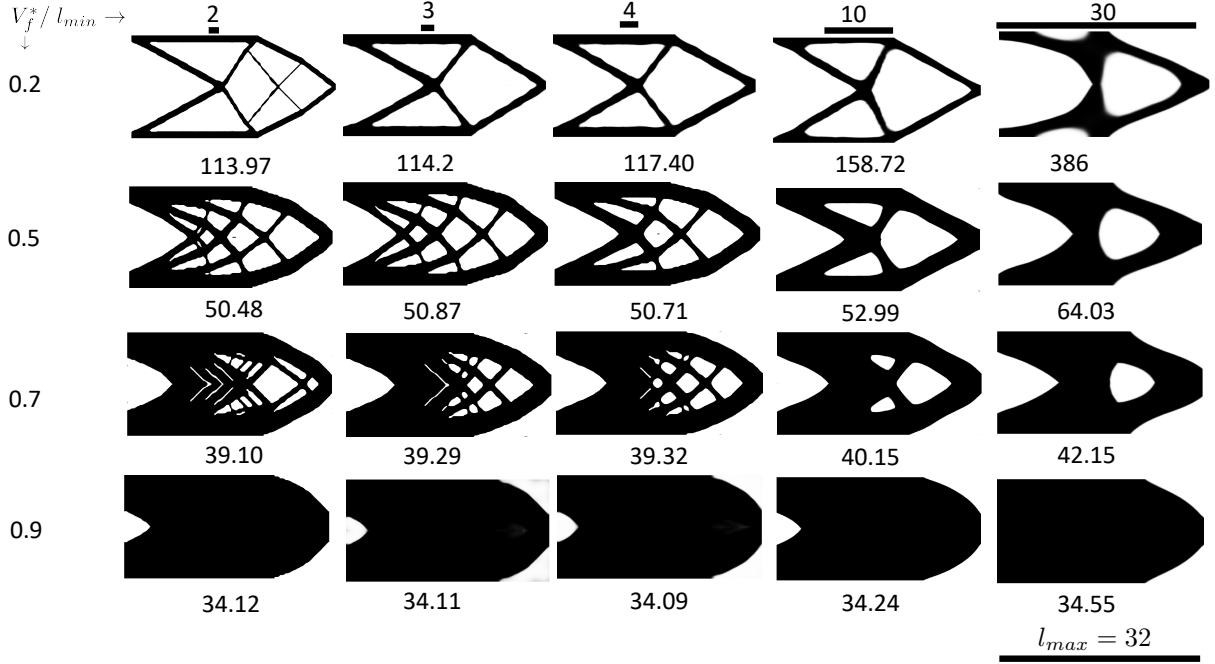}%
		\caption{Varying minimum length scale with $l_{max} = 60$. }
		\label{fig:minLengthScale}
	\end{center}
\end{figure}

Next, we  consider again the mid-cantilever beam in Figure  \ref {fig:benchmark}, and vary $l_{min}$ between 3 and 55, while keeping $l_{max}$ constant at $l_{max} = 60$. The topologies obtained using Fourier-TOuNN for  various volume fractions are illustrated in Figure  \ref{fig:minLengthScale}. Observe that for a given volume fraction, i.e. for each row, as $l_{min}$ is increased, the topology exhibits thicker features. Further  the compliance increases (gets worse) as $l_{min}$ approaches $l_{max}$, i.e., as the solution space shrinks.

Note that a small volume fraction (first row) contradicts the large value of minimum length scale (column 4). The minimum length scale is essentially ignored by the algorithm, while the volume constraint is respected. Such contradictions can be hard to resolve in exact length scale control.

\subsection{Maximum Length-Scale Control}
\label{sec:expts_maxLengthScale}

\begin{figure}
	\begin{center}
		\includegraphics[scale=0.65,trim={70 120 0 0},clip]{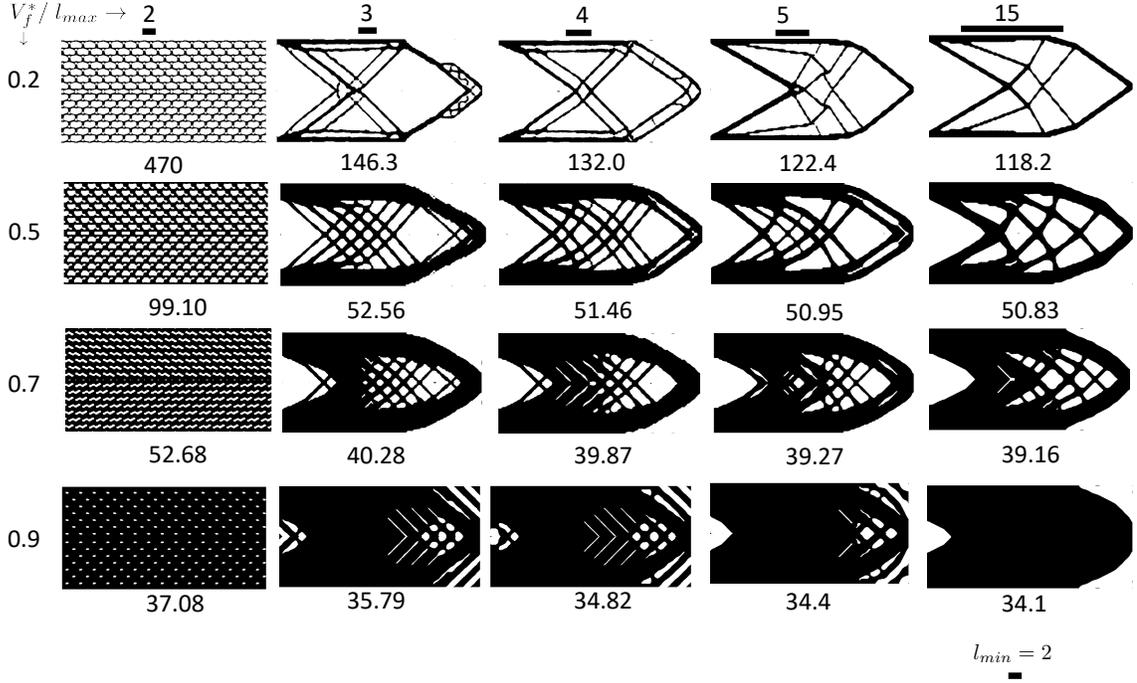}%
		\caption{Varying maximum length scale with $l_{min} = 4$.  }
		\label{fig:maxLengthScale}
	\end{center}
\end{figure}

Next, we  vary $l_{max}$ between 4 and 30, while keeping $l_{min}$ constant at $l_{min} = 4$. The topologies obtained using Fourier-TOuNN for the mid-cantilever beam are illustrated in Figure  \ref{fig:maxLengthScale} for various volume fractions. Similar to the previous experiment, for a given volume fraction, i.e., for each row, as $l_{max}$ is increased, the topologies exhibit thicker features.  Further,  the compliance decreases (improves) as $l_{max}$ moves away from $l_{min}$, i.e., as the solution space expands. 

Once again, note that the large volume fraction imposed (last row)  contradicts the small value of maximum length scale; in the current scenario, this results in disconnected topologies.

\subsection{Single Length-Scale Control}
\label{sec:expts_repeatedPattern}

\begin{figure}
	\begin{center}
		\includegraphics[scale=0.55,trim={40 410 30 70},clip]{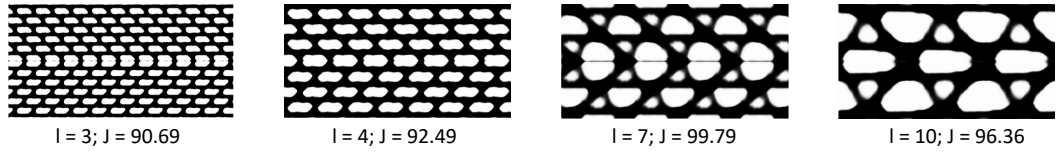}%
		\caption{Repeated patterns for a mid-cantilever beam ($V_f = 0.45$) when $l_{min} = l_{max}$.  }
		\label{fig:narrowBandMidCantilever}
	\end{center}
\end{figure}

In the previous experiment, one can observe that, in  the first column of Figure  \ref{fig:maxLengthScale}, when $l_{min} = l_{max} = 4$, the topologies exhibit a repeated pattern. Indeed, one can obtain repeated patterns of various length scales by setting $l_{min} = l_{max} = l $. This is illustrated in Figure \ref{fig:narrowBandMidCantilever}. Such patterns are often used as fillers in additive manufacturing \cite{Gao2015Review, liu2018current}, and have been obtained by other researchers using various constraint techniques \cite{Wu2017}, \cite{Wu2017ShellInfill}, \cite{wu2016self}. \textcolor{black}{Consistent with the observations of Section \ref{sec:expts_minLengthScale} and \ref{sec:expts_maxLengthScale}, the imposed  length scales can limit the design space, leading to   sub-optimal designs. }

\subsection{Number of Frequency Samples }
\label{sec:expts_freqTerms}

\begin{figure}
	\begin{center}
		\includegraphics[scale=0.35,trim={0 0 0 0},clip]{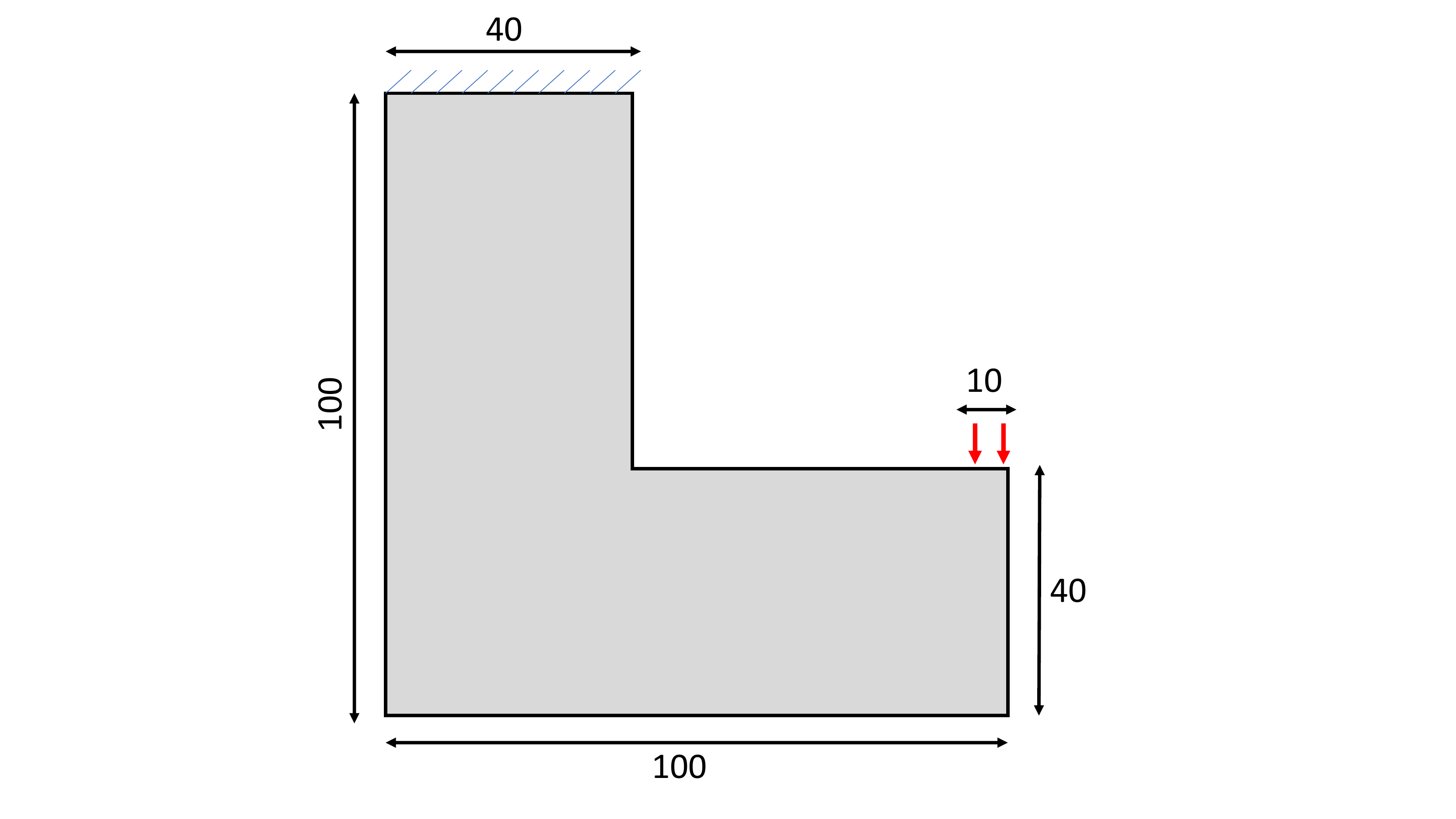}%
		\caption{L-Bracket with load. }
		\label{fig:LbracketDomain}
	\end{center}
\end{figure}

Thus far we have used a default of 150 frequency terms ($n_f$) within the range defined by $l_{min}$ and $l_{max}$. In this experiment, we will vary $n_f$ and study its effect on the topology. In particular, we  consider the L-bracket problem illustrated in Figure \ref{fig:LbracketDomain}. The length scales were kept constant at  $l_{min} = 4$ and  $l_{max} = 30$, and the desired volume fraction is 0.4.

\begin{figure}
	\begin{center}
		\includegraphics[scale=0.75,trim={0 0 0 0},clip]{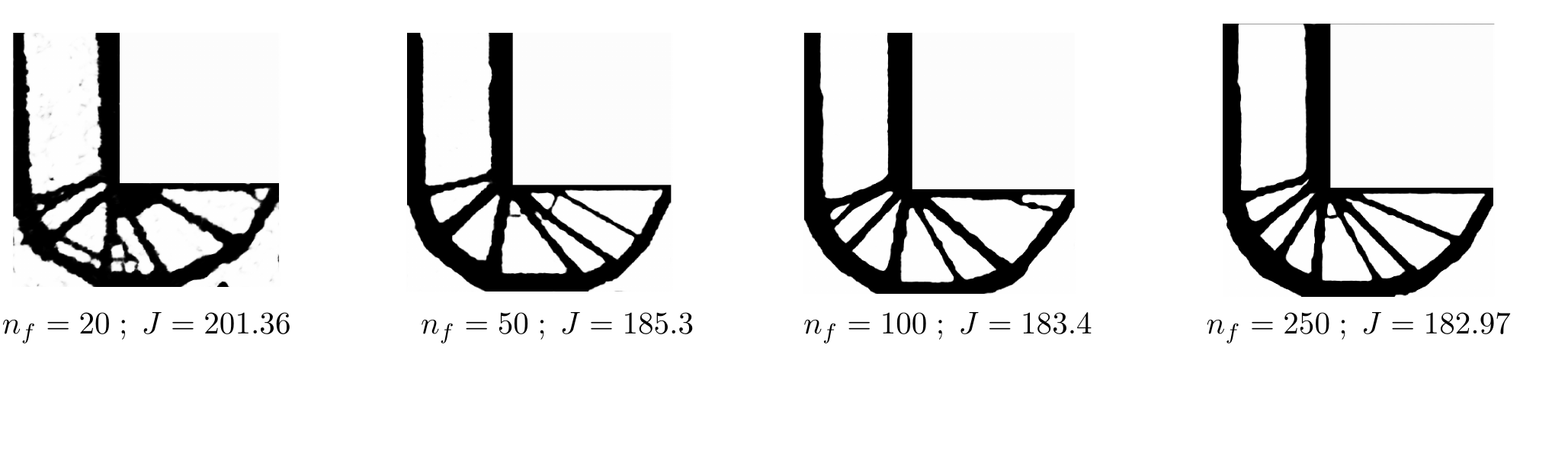}%
		\caption{Effect of number of frequency terms $n_f$ on the topology. }
		\label{fig:numFreqTerms}
	\end{center}
\end{figure}

The  topologies obtained by varying $n_f$ are illustrated in  Figure \ref{fig:numFreqTerms}.   Based on all the experiments reported in this paper, we  observed that about 100 frequency terms are sufficient for convergence. \textcolor {black} {Using very few frequency terms can lead to poor convergence. This is consistent with the observation made in \cite{white2018TOFourier} \emph {" ... if too few Fourier [coefficients] are used, the resulting design is not acceptable."}} 

\subsection{Depth of Neural Net}
\label{sec:expts_NNDepth}

\begin{figure}
	\begin{center}
		\includegraphics[scale=0.75,trim={105 370 100 100},clip]{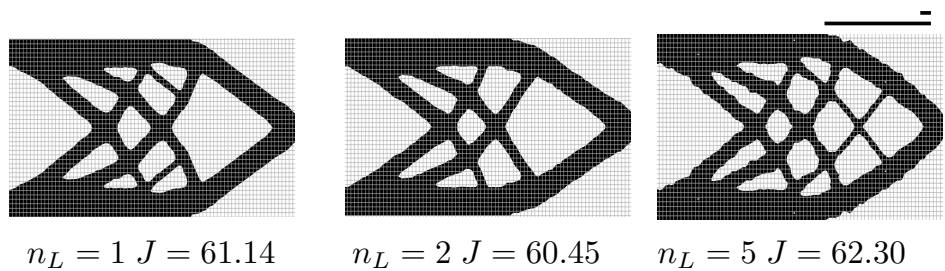}%
		\caption{Effect of the NN's depth on the topology;  the number of design variables are 6040, 6460,  and 7720 respectively.}
		\label{fig:NNDepth}
	\end{center}
\end{figure}

In all the experiments thus far, we have used, and recommend, a NN with just one layer. Here, we briefly consider the effect of the number of layers $n_L$ on the topologies for the mid-cantilever beam with $l_{min} = 2 \; l_{max} = 20$ and $V_f^* = 0.5$.   From Figure \ref {fig:NNDepth}, we observe that as we increase the depth of the NN,  the topologies exhibit wiggles \cite{white2018TOFourier}. \textcolor{black}{This observation is consistent with  analysis  in \cite{tancik2020fourierNN}, where it was shown that  additional layers can lead to frequency spreading. Thus, to limit this spread, we recommend that $n_L = 1$, or $n_L = 2$ in the proposed framework. }

\subsection{\textcolor{black}{Multi-material design}}
\label{sec:expts_MMTO}
\textcolor{black}{ In this section, we demonstrate multi-material design with length scale control. Note that a mass fraction constraint must be imposed, instead of volume fraction. Therefore the physical density of all materials ($[kg \; m^{-3}]$ in $SI$) is required. For the experiments, the candidate materials are listed in Table \ref{table:candidateMaterials}, where  $\lambda$ denotes the physical density.  Observe that. for a given density distribution $\rho^{(i)}(x,y)$,  the total mass is given by:
$$m = \sum\limits_{i=0}^S\int\limits_{\Omega}\lambda^{(i)}\rho^{(i)}(x,y) d\Omega$$
 }
\begin{table}
\centering
\begin{tabular}{|l|l|l|l|}
\hline
Index    & Color & $E$  & $\lambda$ \\ \hline
$\phi$ & Gray  & 1e-9 & 1e-9  \\ \hline
1      & Black & 1    & 1      \\ \hline
2      & Red   & 0.8  & 0.7    \\ \hline
3      & Blue   & 0.2  & 0.15    \\ \hline
\end{tabular}
\caption{Candidate materials}
\label{table:candidateMaterials}
\end{table}
\textcolor{black}{For numerical demonstration, we consider again the Michell structure in Fig. \ref{fig:benchmark}, and impose a desired mass fraction of $0.3$, where a mass fraction of 1 means filling the entire domain with the heaviest material. The mesh size is $60 \times 30$ and  $n_f = 150$.}

\begin{figure}
	\begin{center}
		\includegraphics[scale=0.75,trim={35 30 0 0},clip]{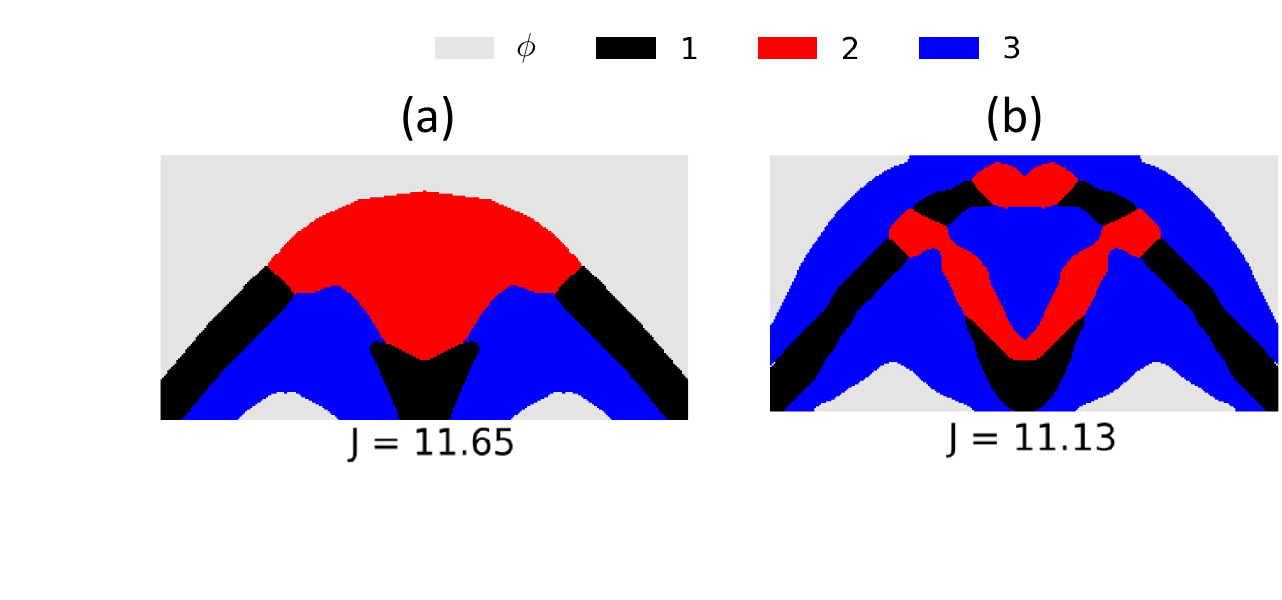}%
		\caption{Three multi-material designs (a) without length scale control \cite{ChandrasekharMMTOuNN2020} (b) with, length scale control.}
		\label{fig:MMTOFourier_Michell}
	\end{center}
\end{figure}

\textcolor{black}{The first experiment involves only material $\phi, 1, 2$ from Table \ref{table:candidateMaterials}. Figure \ref{fig:MMTOFourier_Michell} compares the computed topologies obtained using the original method (without length scale control) \cite{ChandrasekharMMTOuNN2020}, and using the proposed framework, with $l_{min}$ = 3 and $l_{max}$ = 30. Observe that the compliance decreases with length-scale control. Compared to the single material design in Figure \ref{fig:benchmark}, the number of design variables increases marginally from 6040  to 6083.}

\begin{figure}
	\begin{center}
		\includegraphics[scale=1,trim={35 30 0 0},clip]{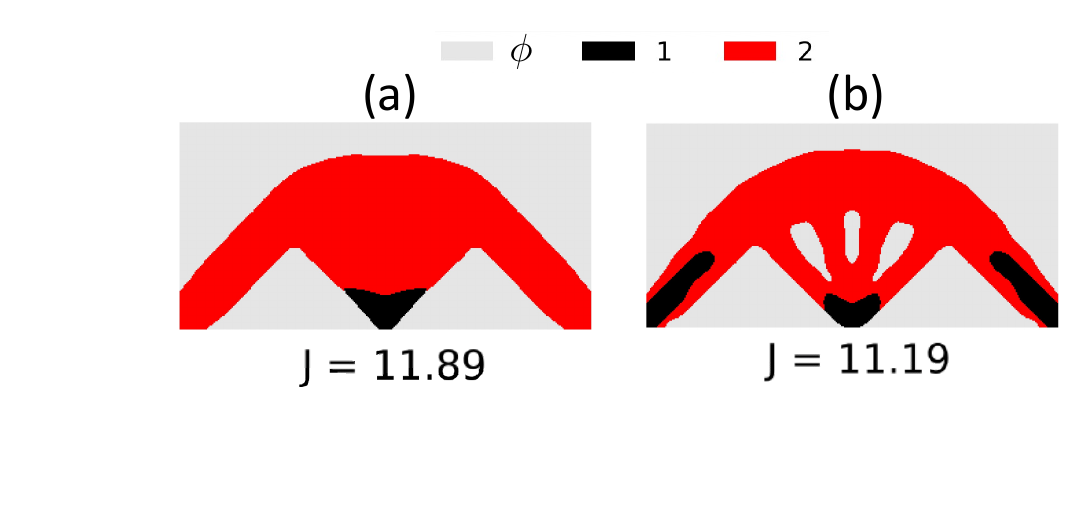}%
		\caption{Two multi-material designs (a) without length scale control \cite{ChandrasekharMMTOuNN2020} (b) with, length scale control.}
		\label{fig:MMTOFourier_Michell_3Mat}
	\end{center}
\end{figure}

\textcolor{black}{Next, we repeat the experiment using all three materials $\phi, 1,2,3$ from Table \ref{table:candidateMaterials}. The topologies  are illustrated  in Fig. \ref{fig:MMTOFourier_Michell_3Mat}. Observe that the volume fraction is fairly high since the light material occupies a significant portion of the design space, but the mass fraction remains at 0.3. The number of design variables increases marginally to 6104.}

\section{Conclusion}
\label{sec:conclusion}

In this paper, we presented a simple length scale control strategy by extending the recently proposed TOuNN method \cite{ChandrasekharTOuNN2020}. The extension relied on a  Fourier projection; no additional constraints were required.  Further, since the computations and  sensitivity analysis are performed via computational graphs and back-propagation,  it is easier to compound other manufacturing constraints \cite{Vatanabe2016ManufConstraints} into the framework.   The optimization was performed using an Adam optimizer. Other schemes such as L-BFGS \cite{Schmidt2020} may lead to faster convergence and needs further experimentation. \textcolor{black}{Further, the FE solver was outside the automatic differentiation framework. Integrating a solver for end-to-end differentiation \cite{chandrasekhar2021auto} will be addressed in the future.} Several numerical experiments were presented to characterize the proposed method. \textcolor{black}{The extension to  multi-material design was also presented.}

\textcolor{black}{The proposed framework has several deficiencies. It only offers an approximate control over the length scales. While this may suffice certain applications, exact length scale control remains to be addressed. As observed earlier, the cost of weight updates also contributes significantly to the net computational cost. Its mitigation needs to be explored. In theory, the method can be extended to 3D, but needs to be demonstrated.}

\section*{Replication of results}
The Python code pertinent to this paper is available at \href{https://github.com/UW-ERSL/Fourier-TOuNN}{https://github.com/UW-ERSL/Fourier-TOuNN}.

\section*{Acknowledgments}
The authors would like to thank the support of National Science Foundation through grant CMMI 1561899.

\section*{Compliance with ethical standards}
The authors declare that they have no conflict of interest.

\bibliographystyle{unsrt}  
\bibliography{AuTO,ERSLReferences,TOuNN_biblio} 
\end{document}